\documentclass[12pt]{article}
\begin{document}
\title{ABOUT GRAVITATION}
\author{B.G. Sidharth\\
International Institute for Applicable Mathematics \& Information Sciences\\
Hyderabad (India) \& Udine (Italy)\\
B.M. Birla Science Centre, Adarsh Nagar, Hyderabad - 500 063 (India)}
\date{}
\maketitle
\begin{abstract}
We consider different deductions of the mysterious Weinberg formula and show that this leads us back to the model of fluctuational cosmology which correctly predicted in advance, dark energy driven, accelerating universe with a small cosmological constant. All this also provides us with an interpretation of Gravitation as the distributional effect of the residual energy of the universe.
\end{abstract}
\section{Introduction}
We consider a model in which $\sqrt{N}$ particles were created out of the background Quantum Vaccuum in an interval $\tau$ in a Critical Point phase transition \cite{ijmpa,ijtp,newcosmos}. That is we have
\begin{equation}
\dot {N} = \sqrt{N}/\tau\label{e1}
\end{equation}
where $N$ denotes the number of particles and $\tau$ a typical elementary particle Compton time.\\
The above 1997 model correctly predicted a dark energy driven accelerating universe with a small cosmological constant. All of this was confirmed by observation over the next six years \cite{perl,kir,science,cu}. Infact from (\ref{e1}) we have on integration
\begin{equation}
T = \sqrt{N} \tau\label{e2}
\end{equation}
where $T$ is the age of the universe and $N \sim 10^{80}$ is the well known number of elementary particles, typically pions, in the universe. Infact (\ref{e2}) immediately leads to the well known so called Eddington formula
\begin{equation}
R = \sqrt{N} l\label{e3}
\end{equation}
Apart from an accelerating universe with a small cosmological constant, the above model also deduced from theory the so called Dirac large number relations and also (\ref{e3}), all of which were known purely as empirical relations without any theoretical basis. The model also deduced the mysterious empirical Weinberg formula
\begin{equation}
m = \left(\frac{H \hbar^2}{Gc}\right)^{\frac{1}{3}}\label{e4}
\end{equation}
where $m$ is the pion mass, $H$ is the Hubble constant and the other symbols having their usual significance. One of the relations deduced is the following:
\begin{equation}
G = \frac{lc^2}{\sqrt{N} m}\label{e5}
\end{equation}
We will now examine the implications of all this.
\section{Characterizing $G$}
We will now digress a little to arrive at the Weinberg formula (\ref{e4}) from an alternative route following Sivaram and Landsberg \cite{sivaram,land}. He uses the gravitational self energy of an elementary particle like the pion with the Uncertainty relation to deduce 
\begin{equation}
\left(\frac{Gm^3c}{\hbar}\right) (T) \approx \hbar\label{e6}
\end{equation}
We use the fact that $T = \frac{1}{H}$ in (\ref{e6}), then we recover the Weinberg formula (\ref{e4}).\\
But if we continue with (\ref{e6}), then we get
\begin{equation}
G = \frac{\hbar^2}{m^3c} \cdot \frac{1}{T}\label{e7}
\end{equation}
The right side of (\ref{e7}) can give back the equation (\ref{e5}). We note that (\ref{e7}) can be rewritten as
\begin{equation}
G = \frac{\hbar^2}{m^3R}\label{ea}
\end{equation}
If in (\ref{ea}) we use the well known relations,
\begin{equation}
R = \frac{GM}{c^2}, M = Nm\label{eb}
\end{equation}
we can get back (\ref{e5}). It must be observed that in the above derivation we have independently obtained (\ref{e2}). Another point to be stressed is that (\ref{e5}) gives a different interpretation of $G$. Rather than being fundamental, it turns out to be ``distributional'' over the whol4e universe (Cf.Section 4). It is the residual energy between the elementary particles over the whole universe, as a result of the Planck scale underpinning of the universe.\\
To continue, we observe that (\ref{e7}) implies that $G$, as in the Dirac cosmology depends on time and is given by
\begin{equation}
G = G_0/T \equiv \frac{l^2c}{m} \cdot \frac{1}{T}\label{e8}
\end{equation}
This time variation given in (\ref{e8}) can be shown to explain standard observations like the precession of the perihelion of Mercury, the bending of light, the decrease in the orbital period of binary pulsars and even the otherwise inexplicable acceleration of the Pioneer spacecrafts \cite{nc,myst}. Equally, we could argue that from (\ref{e8}) we could recover (\ref{e2}) and (\ref{e1}).\\
There is another way to see this. We use the fact that (Cf. Section 4) $n (\sim 10^{40})$ Planck oscillators form the underpinning of a typical elementary particle like the pion which is the ground state. That is,
\begin{equation}
\sqrt{n} mc^2 = m_P c^2\label{e9}
\end{equation}
where $m_P$ is a typical Planck mass $\sim 10^{-5}gms$. The fluctuations in the ground state of this $n$ oscillator system is given by
$\Delta E$ given by (\ref{e9}). So the time for this Uncertainty is given by
\begin{equation}
\hbar / \Delta E = \tau_P\label{e10}
\end{equation}
where $\tau_P$ is the Planck time. In other words there are $\sqrt{n}$ particles due to the uncertainty of fluctuation in the time $\tau_P$. Whence, the rate of the fluctuational appearance of these particles is given by
\begin{equation}
\frac{\sqrt{n}}{\tau_P} = \frac{\sqrt{N}}{\tau}\label{e11}
\end{equation}
(\ref{e11}) is another form of (\ref{e1}).\\
In other words, we arrive at the starting point of Section 1 and then the subsequent relations, from an alternative view point.\\
Finally, it may be observed that the ground state of the $10^{120}$ Planck oscillators which form an underpinning for the universe (Cf.Section 4) has the mass, $10^{-65}gm$. It is quite remarkable that this is the same mass, which according to Landsberg \cite{land}, is the minimum mass observable in the life time of the universe.
\section{The Bi-Scalar Universe}
We now make a few observations. The random walk relations (\ref{e3}) could be rewritten as, at the Planck scale as,
\begin{equation}
l^2 = n l^2_P\label{ea1}
\end{equation}
Equation (\ref{ea1}) brings out the irreducible Quantum of area at the Planck scale, as has proved to be very useful and meaningful in recent Quantum Gravity approaches \cite{baez}. 
It is interesting that we have obtained a rationale for the Quantum of area in terms of a random process. We could think of this in the following model. Let us consider the normals to the Planck area. In the Quantum vaccuum they would be randomly distributed, as in the Ising model \cite{good}. At this stage of incoherence we cannot think of a particle like the pion, which infact as shown above and earlier \cite{psu} is a normal mode of these Planck oscillators. However, let us consider a phase transition at the critical point for these $n$ oscillators, exactly as in Ising model. Infact from critical point theory we have \cite {good,davis,wilson},
\begin{equation}
\bar Q^\nu = \bar \xi^\beta\label{ea2}
\end{equation}
In (\ref{ea2}) $\bar Q$ is the reduced order parameter, the ratio of the constituents in the state of order or coherence to the total number of constituents, while $\bar \xi$ is the reduced coherence length, that is the ratio of the minimum elemental areas in our case $l^2_P$ to the area $l^2$ that emerges due to the critical point transition. That is we have
\begin{equation}
\bar Q \sim \frac{1}{\sqrt{n}}, \bar \xi = (l_P/l)^2,\label{ea3}
\end{equation}
Further from the universality of the exponents at the critical point (Cf.refs.\cite{good,davis}), we have 
$$\nu \approx 2\beta$$
whence we get from (\ref{ea2}) and (\ref{ea3}) the equation
$$l = \sqrt{n} l_P$$
which is nothing but (\ref{e1}) or the area equation (\ref{ea1}).\\
In a similar fashion, working with $l$ and $R$ instead of $l_P, l$ and using (\ref{ea2}) and (\ref{ea3}), we can recover the relation
$$R = \sqrt{N}l,$$
that is (\ref{e3}), as indeed is expected in Critical Point theory.\\
To complete the analysis of this model, we have to consider the behavior of the coupling constants. We know from Critical Point theory that for the coupling constants we have
$$J^{(1)}/kT^{(1)}_c = 1 \, J^{(2)}/kT^{(2)}_c = 1,$$
where in our case, in this model,
$$T^{(1)}_c/T^{(2)}_c = l/R$$
Thus we get
\begin{equation}
J^{(1)}/J^{(2)} = l/R\label{ea4}
\end{equation}
As $J^{(1)} = Gm^2$ and $J^{(2)} = e^2$ are the coupling constants at the scale of gravitation in the macro universe and electromagnetism in the micro universe, we get from (\ref{ea4}) on using (\ref{e2})
\begin{equation}
\frac{Gm^2}{e^2} = \frac{l}{R} = \frac{1}{\sqrt{N}}\label{ea5}
\end{equation}
If our above model is correct (\ref{ea5}) should be valid. Indeed it is a  well known empirical relation  and validates the above model.\\
Interestingly we can derive the result (\ref{ea5}) directly from the Planck scale considerations (instead of the Compton scale derivation).\\
Yet another derivation is from (\ref{e7}). It can be written as
$$Gm^2 = \hbar c \cdot \frac{\hbar}{mc^2} \cdot \frac{1}{T}$$
which gives (\ref{ea5}) if we remember that $\hbar c \sim e^2$ and on using (\ref{e2}).\\
This demonstrates how incoherent Planck oscillators can in phase transition like phenomena form the physical particles of the universe. Similar arguments would apply for elementary particles leading to equation (\ref{e3}).
\section{The Planck, the Compton and the Hubble Scales}
There are a few relations involving microphysical constants and large scale parameters which have been longstanding puzzles. One of them is the relation between the mass of the pion, a typical elementary particle, and the Hubble constant seen earlier, viz. \cite{Weinberg}
\begin{equation}
m \approx \left(\frac{H\hbar^2}{Gc}\right)^{\frac{1}{3}}\label{e1a}
\end{equation}
Another is \cite{Hayakawa,Nottale}
\begin{equation}
R \approx \frac{GM}{c^2}\label{e2a}
\end{equation}
In (\ref{e1a}) $m$ is the pion mass, $H$ the Hubble cosntant, $\hbar$ the reduced Planck constant, $G$ the gravitational constant, $c$ the velocity of light while in (\ref{e2a}) $R$ is the radius of the universe and $M$ its mass. This apart there are the well known so called large number coincidences, made famous by Dirac, involving the number of particles in the universe \cite{cu,Narlikar}. In these and other such relations, the equality is always taken to be in the order of magnitude sense. Further, we follow Dirac and Melnikov in taking constants like $G, c, m$ etc. as being microphysical constants (Cf.ref.\cite{cu}).\\
Weinberg noted that the relation (\ref{e1a}) is inexplicable as it connects constants from microphysics to the cosmological parameter $H$. On the other hand the relation (\ref{e2a}) exhibits the universe as a Schwarzschild black hole (Cf. also \cite{ijtp}. Infact one can even show that the time taken by light to traverse the distance $R$, viz. $T$ the age of the universe is the same as for the interior of a Schwarzschild black hole.)\\
We will now argue that relations like (\ref{e1a}) and (\ref{e2a}) can be explained in terms of a Planck scale underpinning for the universe \cite{psp,psu}.\\
Our starting point is the following observation: The position operator for the Klein-Gordan equation is given by \cite{Schweber}
\begin{equation}
\vec X_{op} = \vec x_{op} - \frac{\imath \hbar c^2}{2} \frac{\vec p}{E^2}\label{e3a}
\end{equation}
From (\ref{e3a}) we get \cite{cu,ijmpa}
\begin{equation}
\hat X^2_{op} \equiv \frac{2m^3 c^4}{\hbar^2} X^2_{op} = \frac{2m^3 c^6}{\hbar^2} x^2 + \frac{p^2}{2m}\label{e4a}
\end{equation}
Mathematically (\ref{e4a}) shows that $\hat X^2_{op}$ gives a problem identical to the harmonic oscillator with quantized levels: Infact the quantized ``space-levels'' for $\vec X^2_{op}$ turn out to be, as can be easily verified, multiples of $(\hbar /mc)^2$!\\
From here, we get $\Delta t = \frac{\Delta x}{c} = \frac{\hbar}{mc^2}$. (A similar analysis can be carried out for the Dirac equation also).\\
As is well known the minimum value for $\hbar/mc$ is the Planck length, which happens to be the Compton length of a Planck mass particle $m_P \sim 10^{-5}gms$, as also its Schwarzschild radius 
$$l_P = Gm_P/c^2$$ 
The minimum Planck length ofcourse plays a crucial role both in Quantum Super Strings theory and Quantum Gravity \cite{Garay,Amati,BGSMUP}. In the modern view, these Planck oscillators are created out of the Quantum Vaccuum.\\
Returning to the mathematical Harmonic oscillator problem for the Klein-Gordon (spinless) equation, as is well known \cite{BD2},the quantized energy (in our case length squared) levels are given as multiples of the number of oscillators\ $N$ and the minimum (or ``ground'') state:
\begin{equation}
(\mbox{Length})^2 \sim Nl^2\label{e5a}
\end{equation}
$l$ being the Planck /Compton length in question.\\
We thus have the following equations:
\begin{equation}
R = \sqrt{N} l_P =  \sqrt{\bar N} l, \, l = \sqrt{n} l_P\label{e6a}
\end{equation}
where $N, \bar N$ and $n$ are certain large numbers whose value we will get below and $l$ is the Compton wavelength of a typical elementary particle like the pion. In the second of (\ref{e6a}), we consider the fact that there are $\bar N$ elementary particles in the universe, while in the third relation, an elementary particle like the pion has an underpinning of $n$ Planck oscillators. \\
From (\ref{e6a}) it can easily be seen that
\begin{equation}
N = \bar N n\label{e7a}
\end{equation}
If in (\ref{e6a}) we use the explicit expression for the Compton length, we can easily deduce that, $m$ being the pion mass,
\begin{equation}
m_P = \sqrt{n} m\label{e8a}
\end{equation}
As we have supposed that there are $N$ Planck masses in the universe, their gravitational energy (they do not have any electromagnetic energy as they are chargeless) should equal the energy of the universe that is
$$\frac{GNm^2_P}{R} = Mc^2,$$
$M$ being the mass of the universe.\\
Using the fact that the Planck length is also the Schwarschild radius of the Planck mass we get with (\ref{e6a}),
\begin{equation}
M = \sqrt{N} m_P\label{e9a}
\end{equation}
Similarly if we use the fact that $n$ Planck masses make up the pion, we have
$$\frac{Gnm^2_P}{R} = mc^2$$
Whence we get
\begin{equation}
M = \sqrt{Nn} m\label{e10a}
\end{equation}
The above equations can also be written as 
$$\frac{l_P m_P n}{R} = m$$
Whence it also follows that
\begin{equation}
n = \sqrt{\bar N}\label{e11a}
\end{equation}
So,
\begin{equation}
N = \bar N^{3/2}\label{e12a}
\end{equation}
And
\begin{equation}
M = \bar N m\label{e13a}
\end{equation}
If we now use the fact that the entire gravitational energy of the pions gives its energy, that is, using (\ref{e13a}),
$$\frac{G\bar N m^2}{R} =  mc,^2$$
we will get
\begin{equation}
R = \frac{GM}{c^2}\label{e14a}
\end{equation}
It will be seen that (\ref{e14a}) is the same as (\ref{e2a}). Infact we could obtain (\ref{e14a}) alternatively from (\ref{e9a}) and its preceding equation.\\
If we use (\ref{e6a}) and subsequent equations in (\ref{e14a}) we get immediately
\begin{equation}
G = \frac{c^2l}{m \sqrt{\bar N}}\label{e15a}
\end{equation}
Whence we have
\begin{equation}
\bar N = \left\{\frac{c^2 l}{mG}\right\}^2 \sim 10^{80}\label{e16a}
\end{equation}
We have thus deduced the well known number of elementary particles (like the pions) in the universe, in terms of the microphysical constants in (\ref{e16a}). Immediately we can see that from equations like (\ref{e7a}), (\ref{e11a})  and (\ref{e12a}) that $N \sim 10^{120}$ and $n \sim 10^{40}$. It is now also possible to deduce the mysterious Weinberg formula (\ref{e1a}) from these equations. Infact this follows from (\ref{e1a}) and (\ref{e6a}) if we remember that
$$H = c/R$$
where $H$ is the Hubble constant. Moreover, it can be seen that equations like (\ref{e8a}), (\ref{e9a}) and (\ref{e10a}) are all consistent.\\
Using the fact that from the Klein-Gordan equation, the squares of the length intervals are quantized, equations (\ref{e3a}), (\ref{e4a}) and (\ref{e5a}) and using the fact that the minimum length is at the Planck scale, which therefore provides an underpinning for the universe, and using only the values of the microphysical constants, all of the hitherto mysterious empirical large number relations and the Weinberg formula are seen to be a consequence of the theory and are deducable. They are not adhoc empirical relations as hitherto supposed.
\section{Remarks}
1. We observe that (\ref{e14a}) also be obtained in a Friedman uniformly expanding universe. Given (\ref{e14a}), using (\ref{e6a}), we can deduce (\ref{e9a}), (\ref{e11a}) and (\ref{e12a}) and thence (\ref{e15a}) (0r (\ref{e16a})). From (\ref{e15a}) we have
$$\dot {G} = - \frac{1}{2} N^{-3/2} \cdot (G_0/\tau) (\dot {N}) = G_0/T^2$$
\begin{equation}
\mbox{or} \, + \frac{1}{2} N^{-1/2} \cdot \tau \frac{G_0}{T^2} (\dot {N}) = \frac{G_0}{T^2} = \dot {N} = \frac{\sqrt{N}}{\tau}\label{e17a}
\end{equation}
That is we recover (\ref{e17a}). This is an alternative route to fluctuational cosmology.\\
2. In the context of self energy, we can get a clue to the origin of the energy of a particle in the Dirac development in the following manner. Following Dirac \cite{dirac}, we have
\begin{equation}
\imath \hbar \ddot{\alpha} = 2 \dot {\alpha} H\label{e18a}
\end{equation}
where $c \alpha$ is the velocity operator, and
\begin{equation}
\dot {\alpha} = \dot {\alpha}  e^{-2\imath H t/\hbar}\label{e19a}
\end{equation}
Whence we can deduce
\begin{equation}
x = - \frac{1}{4} \imath \hbar^2 \dot {\alpha} (0) \left(e^{-2\imath H t/\hbar}\right)H^{-2}\label{e20a})
\end{equation}
Combining (\ref{e20a}) with (\ref{e19a}) we get
\begin{equation}
\ddot {x} = - \frac{H^2}{c\hbar^2} x\label{e21a}
\end{equation}
Equation (\ref{e21a}) shows that, neglecting the external motion of the particle, within the Compton wavelength the particle behaves like the Harmonic oscillator. We can get a better idea of this oscillator system if we consider the classical equation 
\begin{equation}
m  \ddot{x} = F_{ext} (\dot {x}) + \frac{2}{3} \left(e^2/4\pi c^3\right) \frac{d}{dt} (\ddot{x})\label{e22a}
\end{equation}
In the absence of any external force we have the self force given by
\begin{equation}
k_{\mbox{self}} \equiv \frac{2}{3} \frac{e^2}{4\pi c^3} \frac{d}{dt} (\ddot{x})\label{e23a}
\end{equation}
On the one hand, (\ref{e23a}) can be identified with (\ref{e18a}) on using (\ref{e19a}) and (\ref{e20a}) and all this leads us back to the Harmonic oscillator equation (\ref{e18a}) which gives the energy levels for the particle. This can be identified with the inertial energy of the particle.\\
3. So called extremal black holes were discussed previously \cite{bgskluwer}. These are charged black holes with charge given by the relation
\begin{equation}
Q \sim Mc^2\label{e24a}
\end{equation}
Interestingly these very short lived particles as was pointed out would soon annihilate producing gamma rays. Such $MeV$ particles have been proposed recently \cite{boehm} as candidates for the gamma rays. On the other hand, it was also pointed out that with masses $\sim 10^{-8}$ of  the electron mass we get the   neutrino mass with a  charge that would be one millionth that of the electron and interestingly this gives the correct weak interaction strength.

\end{document}